\documentstyle[12pt]{article}

\catcode`\@=11
\long\def\@makefntext#1{
\protect\noindent \hbox to 3.2pt {\hskip-.9pt
$^{{\ninerm\@thefnmark}}$\hfil}#1\hfill}		

\def\@makefnmark{\hbox to 0pt{$^{\@thefnmark}$\hss}}  

\def\ps@myheadings{\let\@mkboth\@gobbletwo
\def\@oddhead{\hbox{}
\rightmark\hfil\ninerm\thepage}
\def\@oddfoot{}\def\@evenhead{\ninerm\thepage\hfil
\leftmark\hbox{}}\def\@evenfoot{}
\def\sectionmark##1{}\def\subsectionmark##1{}}

\renewcommand{\thefootnote}{\fnsymbol{footnote}}

\newcounter{sectionc}\newcounter{subsectionc}\newcounter{subsubsectionc}
\renewcommand{\section}[1] {\vspace*{0.6cm}\addtocounter{sectionc}{1}
\setcounter{subsectionc}{0}\setcounter{subsubsectionc}{0}\noindent
	{\normalsize\bf\thesectionc. #1}\par\vspace*{0.4cm}}
\renewcommand{\subsection}[1] {\vspace*{0.6cm}\addtocounter{subsectionc}{1}
	\setcounter{subsubsectionc}{0}\noindent
	{\normalsize\it\thesectionc.\thesubsectionc. #1}\par\vspace*{0.4cm}}
\renewcommand{\subsubsection}[1]
{\vspace*{0.6cm}\addtocounter{subsubsectionc}{1}
	\noindent {\normalsize\rm\thesectionc.\thesubsectionc.\thesubsubsectionc.
	#1}\par\vspace*{0.4cm}}

\newcounter{appendixc}
\newcounter{subappendixc}[appendixc]
\newcounter{subsubappendixc}[subappendixc]

\renewcommand{\appendix}[1] {\vspace*{0.6cm}
        \refstepcounter{appendixc}
        \setcounter{figure}{0}
        \setcounter{table}{0}
        \setcounter{equation}{0}
        \renewcommand{\thefigure}{\Alph{appendixc}.\arabic{figure}}
        \renewcommand{\thetable}{\Alph{appendixc}.\arabic{table}}
        \renewcommand{\theappendixc}{\Alph{appendixc}}
        \renewcommand{\theequation}{\Alph{appendixc}.\arabic{equation}}
        \noindent{\bf Appendix \theappendixc #1}\par\vspace*{0.4cm}}

\def\abstracts#1{{

\centering{\begin{minipage}{12.2truecm}\footnotesize\baselineskip=12pt\noindent
	\centerline{\footnotesize ABSTRACT}\vspace*{0.3cm}
	\parindent=0pt #1
	\end{minipage}}\par}}

\renewenvironment{thebibliography}[1]
	{\begin{list}{\arabic{enumi}.}
	{\usecounter{enumi}\setlength{\parsep}{0pt}
\setlength{\leftmargin 1.25cm}{\rightmargin 0pt}
	 \setlength{\itemsep}{0pt} \settowidth
	{\labelwidth}{#1.}\sloppy}}{\end{list}}

\newcounter{itemlistc}
\newcounter{romanlistc}
\newcounter{alphlistc}
\newcounter{arabiclistc}

\newcommand{\fcaption}[1]{
        \refstepcounter{figure}
        \setbox\@tempboxa = \hbox{\footnotesize Fig.~\thefigure. #1}
        \ifdim \wd\@tempboxa > 6in
           {\begin{center}
        \parbox{6in}{\footnotesize\baselineskip=12pt Fig.~\thefigure. #1}
            \end{center}}
        \else
             {\begin{center}
             {\footnotesize Fig.~\thefigure. #1}
              \end{center}}
        \fi}

\newcommand{\tcaption}[1]{
        \refstepcounter{table}
        \setbox\@tempboxa = \hbox{\footnotesize Table~\thetable. #1}
        \ifdim \wd\@tempboxa > 6in
           {\begin{center}
        \parbox{6in}{\footnotesize\baselineskip=12pt Table~\thetable. #1}
            \end{center}}
        \else
             {\begin{center}
             {\footnotesize Table~\thetable. #1}
              \end{center}}
        \fi}

\def\@citex[#1]#2{\if@filesw\immediate\write\@auxout
	{\string\citation{#2}}\fi
\def\@citea{}\@cite{\@for\@citeb:=#2\do
	{\@citea\def\@citea{,}\@ifundefined
	{b@\@citeb}{{\bf ?}\@warning
	{Citation `\@citeb' on page \thepage \space undefined}}
	{\csname b@\@citeb\endcsname}}}{#1}}

\newif\if@cghi
\def\cite{\@cghitrue\@ifnextchar [{\@tempswatrue
	\@citex}{\@tempswafalse\@citex[]}}
\def\citelow{\@cghifalse\@ifnextchar [{\@tempswatrue
	\@citex}{\@tempswafalse\@citex[]}}
\def\@cite#1#2{{$\null^{#1}$\if@tempswa\typeout
	{IJCGA warning: optional citation argument
	ignored: `#2'} \fi}}

 1
 1
 1

\font\ninerm=cmr9

\begin{document}

\newcommand{\st}{\scriptstyle}
\newcommand{\sst}{\scriptscriptstyle}
\newcommand{\mco}{\multicolumn}
\newcommand{\epp}{\epsilon^{\prime}}
\newcommand{\vep}{\varepsilon}
\newcommand{\ra}{\rightarrow}
\newcommand{\ppg}{\pi^+\pi^-\gamma}
\newcommand{\vp}{{\bf p}}
\newcommand{\ko}{K^0}
\newcommand{\kb}{\bar{K^0}}
\newcommand{\al}{\alpha}
\newcommand{\ab}{\bar{\alpha}}
\def\be{\begin{equation}}
\def\ee{\end{equation}}
\def\bea{\begin{eqnarray}}
\def\eea{\end{eqnarray}}
\def\CPbar{\hbox{{\rm CP}\hskip-1.80em{/}}}
\def\anti{\overline}
\def\rta{\rightarrow}
\def\hsm{\phi^0}
\def\mhsm{m_{\phi^0}}
\def\gev{~{\rm GeV}}
\def\gam{\gamma}
\def\tanb{\tan\beta}
\def\mha{m_A}

\centerline{\normalsize\bf SEARCHING FOR HIGGS BOSONS ON LHC }
\baselineskip=22pt
\centerline{\normalsize\bf USING $b$-TAGGING}

\centerline{\footnotesize Jin Dai}
\baselineskip=13pt
\centerline{\footnotesize\it Dept. of Physics, U. C. San Diego,
    Lo Jolla, CA, 92093-0319 }
\centerline{\footnotesize E-mail: dai@higgs.ucsd.edu}
\vspace*{0.3cm}
\centerline{\footnotesize and}
\vspace*{0.3cm}
\centerline{\footnotesize J.F. Gunion }
\baselineskip=13pt
\centerline{\footnotesize\it Davis Institute for High Energy Physics,
    Dept. of Physics, U.C. Davis, Davis, CA 95616}
\vspace*{0.3cm}
\centerline{\footnotesize and}
\vspace*{0.3cm}
\centerline{\footnotesize R. Vega }
\baselineskip=13pt
\centerline{\footnotesize\it Dept. of Physics, Southern Methodist
           University, Dallas, TX 75275}
\vspace*{0.3cm}
\abstracts{ We demonstrate that the detection of the SM and MSSM
Higgs bosons will be possible at the LHC via $t\anti t b\anti b$ and
$b\anti b b\anti b$ final state, provided $b$-tagging can be performed
with good efficiency and purity. }

\normalsize\baselineskip=15pt
\setcounter{footnote}{0}
\renewcommand{\thefootnote}{\alph{footnote}}
\section{Introduction}
Understanding the Higgs sector is one
of the fundamental missions of future high energy colliders such
as the LHC. However, options for detection of the Standard Model (SM)
Higgs boson, $\hsm$, are limited to rare $\gam\gam$ decay chanel
if $\mhsm$ lies between $ 80 \gev$ and $ 130 \gev$.
The Higgs bosons of the Minimal Supersymmetric Standard Model (MSSM)
is even harder to find. In fact, there may be a window
of $\mha$--$\tanb$ parameter space, in which
no MSSM Higgs boson would be seen either at LEP-II or at the LHC.

Clearly, the establishment of viable
techniques for detection of the Higgs bosons in its main decay mode in the
intermediate mass region, $h\rta b\anti b$, would be highly desirable.
In this talk, we present our works\cite{DGV}
on using $b$-tagging to search for Higgs bonsons.

The basic idea is to look for Higgs bosons produced associated with
$t\anti t$ or $b\anti b$, which then decays into $b\anti b$, hence
the final state is $t\anti t b\anti b$ or $b\anti b b\anti b$. In
either case, there will be four $b$-quarks in the final state. By tagging
three or more $b$'s, QCD background can be suppressed and the detection
of the Higgs mass peak is possible.

\section{MonteCarlo Simultion}

The major background to the process that we discussed is of two types:
1)Inpurity backgrounds: $b\anti b g$ or $t\anti t$, $t\anti t g$
production with one of the gluons or light quarks in the final state
being mistagged as a $b$-jet; 2) Irreducible backgrounds: direct QCD
production of $t\anti t b\anti b$ or $b\anti b b\anti b$. They
are of the same order of magnitude.

When Higgs mass is close to $m_Z$, $t\anti t Z$ background to $t\anti t
h$ are relevant, but Higgs peak will not be confused as a $Z$ peak due
to smaller branching ratio of $Z\rta b\anti b$. Other types of
backgrounds are considered but are negligilbe.

For $t\anti t b\anti b$ final states, we always require one of the top
quarks to decay semileptonically so the event can be easily triggered on
the lepton, we also require tagging three or all four $b$-jets.
For $b\anti b b\anti b$ final states, we simply require tagging three
$b$'s, special triggers may have to be designed.

We did the tree level parton model MonteCarlo simulation of both the
signal and the background with the $K$ factor put in. Semileptonic
$b$ decays, which will change the $b\anti b$ invariant mass distribution
due to missing momenta from the neutrino, is taken into account in our
latest work. This will bring our parton level simulation much closer
to jet level simulation.

For $b\anti b b\anti b$ chanel, the singal/background ratio is very
small in some cases even though the singal is statistically $5\sigma$,
we discussed the procedure needed to recover such a weak Higgs peak from
the overwhelming background and made positive conclusions.

\section{Result}

We will briefly describe our result here and refer to our paper for
detailed description.

We assume 30\% and $b$-tagging efficiency and 1\% of mistagging
probability. (See our paper \cite{DGV} for detailed description of
cuts.) With several years of running at $100fb^{-1}$ per year, the
$t\anti t b\anti b$ chanel can be used to discover the SM Higgs up to
110-120GeV, and can be combined with the $Wbb$ chanel\cite{SMW}. This
is complementary to the $\gam\gam$ chanel, which gets worse when Higgs is
lighter.

With $100fb^{-1}$, at $4\sigma$ level, MSSM Higgs $h$ can be dicovered
for a large part of parameter region and $A$ for moderate $\tanb$
through $t\anti t b\anti b$ chanel. In a large triangle region for
high $\tanb$ and moderate $m_A$, $h$ and $A$ or $H$ and $A$ can be
simualtaneously discovered via $b\anti b b\anti b$ final states. The window
which is mentioned ealier is now covered. In a more optimistic senario,
$200fb^{-1}$ with 40\% $b$-tagging efficiency and same purity, the
above two chanels alone can cover essentially all the MSSM
parameter space, and often several Higgs bonsons can be discovered
via several chanels on LHC.

\section{Conclusion}
The very great promise of these modes
makes it virtually mandatory that the detector collaborations at the
LHC find a way to perform $b$-tagging with the required efficiency
and purity in the multi-event-per-crossing environment that
will prevail for high instantaneous luminosity at the LHC.

\end{document}